\documentclass[aps,prl,twocolumn,showpacs,groupedaddress]{revtex4}  
\usepackage{graphicx}  
\usepackage{dcolumn}   
\usepackage{bm}        
\usepackage{amssymb}   

\newcommand{\BaBar}{\textsc{BaBar }}

\newcommand{\nn}{\nonumber \\}

\newcommand{\ov}[1]{\overline{#1}}

\newcommand{\etal}{\emph{et al.}}
\newcommand{\rf}[1]{Ref.\ \cite{#1}}

\newcommand{\adcpv}{A_{CP}(B^+ \to J/\psi K^+)}

\newcommand{\acp}{A_{CP}}

\newcommand{\jp}{J/\psi}
\newcommand{\jpsi}{J/\psi}

\newcommand{\gevcc}{GeV/$c^2$}

\newcommand{\fig}[1]{Fig.~\ref{#1}}
\newcommand{\tab}[1]{Table \ref{#1}}

\makeatletter
\newcommand\figcaption{\def\@captype{figure}\caption}
\newcommand\tabcaption{\def\@captype{table}\caption}
\makeatother

\begin{document}


\hspace{5.2in} \mbox{FERMILAB-PUB-08-039-E}

\title{Study of direct CP violation in 
$B^{\pm} \to J/\psi K^{\pm}(\pi^{\pm})$ decays}
%
\author{V.M.~Abazov$^{36}$}
\author{B.~Abbott$^{75}$}
\author{M.~Abolins$^{65}$}
\author{B.S.~Acharya$^{29}$}
\author{M.~Adams$^{51}$}
\author{T.~Adams$^{49}$}
\author{E.~Aguilo$^{6}$}
\author{S.H.~Ahn$^{31}$}
\author{M.~Ahsan$^{59}$}
\author{G.D.~Alexeev$^{36}$}
% ...

\affiliation{\vspace{0.1 in}(The D\O\ Collaboration)\vspace{0.1 in}}
% ...
\affiliation{$^{6}$University of Alberta, Edmonton, Alberta, Canada,
                Simon Fraser University, Burnaby, British Columbia, Canada,
                York University, Toronto, Ontario, Canada, and
                McGill University, Montreal, Quebec, Canada}
% ...
\affiliation{$^{29}$Tata Institute of Fundamental Research, Mumbai, India}
% ...
\affiliation{$^{31}$Korea Detector Laboratory, Korea University, Seoul, Korea}
% ...
\affiliation{$^{36}$Joint Institute for Nuclear Research, Dubna, Russia}
% ...
\affiliation{$^{49}$Florida State University, Tallahassee, Florida 32306, USA}
% ...
\affiliation{$^{51}$University of Illinois at Chicago,
                Chicago, Illinois 60607, USA}
% ...
\affiliation{$^{59}$Kansas State University, Manhattan, Kansas 66506, USA}
% ...
\affiliation{$^{65}$Michigan State University,
                East Lansing, Michigan 48824, USA}
% ...
\affiliation{$^{75}$University of Oklahoma, Norman, Oklahoma 73019, USA}
% ...
\date{February 22, 2008}

\begin{abstract}

We present a search for direct CP violation in 
$B^{\pm} \to J/\psi K^{\pm}(\pi^{\pm})$ decays.  
The event sample is selected from 2.8 fb$^{-1}$ 
of $p\ov{p}$ collisions
recorded by D0 experiment in Run II of the
Fermilab Tevatron Collider.
The charge asymmetry 
$A_{CP}(B^+ \to \jp K^+) = +0.0075 \pm 0.0061$(stat.)$\pm 0.0027$(syst.)
is obtained using a sample of approximately 40 thousand
$B^{\pm} \to J/\psi K^{\pm}$ 
decays. The achieved precision is of the same level as the expected deviation 
predicted by some extensions of the standard model.
We also measured the charge asymmetry 
$A_{CP}(B^+ \to \jp \pi^+) = -0.09 \pm 0.08$(stat.)$\pm 0.03$(syst.).
\end{abstract}


\maketitle



This Letter presents a study of the charge asymmetry
in the decay $B^\pm \to \jp K^\pm$($\pi^\pm$), which is defined as
\begin{eqnarray}
\acp(B^+ \to \jp K^+(\pi^+)) = {} \qquad \qquad \qquad \nn
\frac{N(B^- \to \jp K^-(\pi^-)) - N(B^+ \to \jp K^+(\pi^+))}
              {N(B^- \to \jp K^-(\pi^-)) + N(B^+ \to \jp K^+(\pi^+))}.\nonumber
\label{a-dcpv}
\end{eqnarray} 
A non-zero value of $\acp(B^+ \to \jp K^+(\pi^+))$ corresponds to 
direct CP violation in this decay. In the $b \to s c\ov{c}$ transition
(charge conjugate states are assumed throughout),
the tree-level and $b \to s$ penguin amplitudes have 
a small relative weak phase, 
arg$[-V_{cs}V_{cb}^* / V_{ts}V_{tb}^*]$.
Therefore, the standard model (SM) predicts a
small $\acp(B^+ \to \jp K^+) \sim 0.003$ \cite{hou}. 
Thus, the measurement
of $\acp(B^+ \to \jp K^+)$ is an important way of constraining
those new physics models which predict 
an enhanced value
of this asymmetry, up to 0.01 or higher \cite{hou}.
Most cited are the models with an extra $U(1)'$ gauge boson
responsible for the flavor-changing coupling between $b$ and $s$ quarks \cite{zprime}
and the Two-Higgs Doublet Model (2HDM), which introduces 
an extra coupling to the charged Higgs boson~\cite{thdm}.

In $b \to d c\ov{c}$ transitions, 
on the contrary, 
the relative phase between the tree-level 
and $b \to d$ penguin diagram,
arg$[-V_{cd}V_{cb}^* / V_{td}V_{tb}^*]$,
is expected to be significant so that direct CP-violation 
may be of the order of one percent \cite{dunietz,hou2}.
Decays governed by $b \to d c\ov{c}$ transition
have already been explored.
Recently the Belle collaboration reported large direct CP violation
in $B^0 \to D^+ D^-$
decays, $\mathcal{A}_{D^+D^-}=+0.91\pm0.23\pm0.06$ \cite{belle}, 
much in excess of the SM expectation.
However, this result was not confirmed by the \BaBar collaboration, 
which measured 
$C_{D^+D^-}=-\mathcal{A}_{D^+D^-}=+0.11\pm0.22\pm0.07$~\cite{babar}. 
Here, we report a complementary measurement
of the direct CP violation asymmetry in the $b \to d c\ov{c}$ transition
using the decay $B^+ \to \jp \pi^+$.




The D0 detector is described in detail elsewhere~\cite{run2det}. The
detector components most important for this analysis are the central
tracking and muon system. 
The D0 central tracking system consists
of a silicon microstrip tracker (SMT) and a central fiber tracker
(CFT), both located within a 2~T superconducting solenoidal magnet.
The muon system is located outside the
calorimeters and consists
of a layer of tracking detectors and scintillation trigger counters in
front of 1.8~T iron toroids, followed by two similar layers behind the
toroids~\cite{run2muon}.
The polarities of the solenoid and toroid
are reversed regularly during data taking, so that the four solenoid-toroid
polarity combinations are exposed to approximately the same integrated
luminosity. The reversal of magnet polarities
is essential to reduce the detector-related systematics in asymmetry 
measurements and is fully exploited in this study.


The decay chain $B^+ \to \jp K^+(\pi^+)$ with $\jp \to \mu^+ \mu^-$ is selected 
for this analysis from 2.8 fb$^{-1}$ recorded by D0.
Each muon is required to be identified by the muon system,
to have an associated track in
the central tracking system with at least two
measurements in the SMT, and a transverse momentum $p_T^{\mu} > 1.5$ GeV/$c$
with respect to the beam axis. At least one of the two muons is required to
have matching track segments both inside and outside the toroidal magnet.
The di-muon system must have a reconstructed invariant mass 
between 2.80 GeV/$c^2$ and 3.35 GeV/$c^2$.
An additional charged particle with $p_T > 0.5$ GeV/$c$, total momentum
above 0.7 GeV/$c$, and at least two measurements in the SMT, is selected.
This particle is assigned the kaon mass and is required
to have a common vertex with the two muons, 
with the $\chi^2$ of the vertex fit being less than 16 for three degrees of freedom. 
The displacement of this vertex from 
the primary interaction point is required to exceed three standard deviations 
in the plane perpendicular to the beam direction.
The primary vertex of the $p \bar{p}$ interaction is determined for each event
using the method described in Ref.~\cite{btag}.
The average position of the beam-collision point is included as a constraint.


From each set of three particles fulfilling these requirements,
a $B^+$ candidate is constructed.
The momenta of the muons are corrected using the $\jp$ mass constraint.
To further improve the $B^+$ selection, a likelihood ratio method 
\cite{bgv} is applied.
The variables chosen for this analysis
include the lower transverse momentum of the two muons, the $\chi^2$
of the $B^+$ decay vertex fit, the $B^+$ decay length divided by its
uncertainty, the significance $S_B$ of the $B^+$ track impact 
parameter, the transverse momentum of the kaon, 
and the significance $S_K$ of the kaon track
impact parameter. For any track $i$, the significance is defined as
$S_i = \sqrt{[\epsilon_T/\sigma(\epsilon_T)]^2 +
 [\epsilon_L/\sigma(\epsilon_L)]^2}$, where $\epsilon_T$ ($\epsilon_L$)
is the projection of the track impact parameter on the plane
perpendicular to the beam direction (along the beam direction),
and $\sigma(\epsilon_T)$ [$\sigma(\epsilon_L)$] is its uncertainty.
The track of each $B^+$ is fitted assuming that
it passes through the reconstructed vertex and is directed
along the reconstructed $B^+$ momentum.
Finally, the mass of the reconstructed $B^+$ candidate
is constrained to the window $4.98<m(\jp K)<5.76$ \gevcc.



\begin{figure}
\begin{center}
\includegraphics[width=0.45\textwidth]{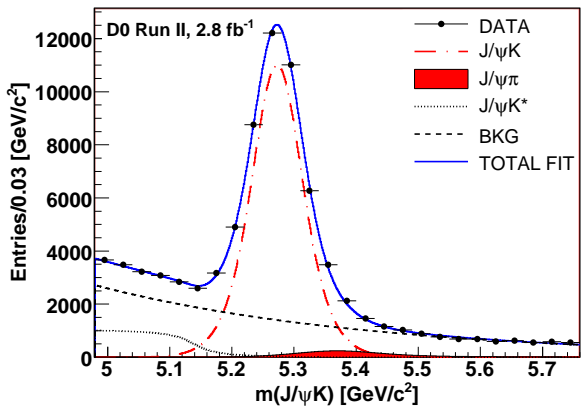}
\caption{The $J/\psi K$ invariant mass distribution together with 
the result from the unbinned likelihood fit (the unsplitted sample).}
\label{mjpk-fit}
\end{center}
\end{figure}

The resulting invariant mass distribution
of the $\jpsi K$ system is shown in \fig{mjpk-fit}
with the result of an unbinned likelihood fit to the sum
of contributions from $B \to \jpsi K$, $B \to \jpsi \pi$, and
$B \to \jpsi K^{*}$ decays, as well as combinatorial background (BKG). 
The mass distribution
of the $\jpsi K$ system from the $B \to \jpsi K$ hypothesis
is parameterized by a Gaussian function with the width depending
on the momentum of the $K$ candidate. The mass
distribution of the $\jpsi \pi$ system
from the $B \to \jpsi \pi$ hypothesis is parameterized
by a Gaussian function with the same width. It is then transformed into
the distribution of the $\jpsi K$ system by assigning the kaon mass
to the pion. 
The decay $B \to \jpsi K^{*}$ with $K^{*} \to K \pi$,
where the pion is not reconstructed, produces a
broad $\jpsi K$ mass distribution with the threshold near $m(B) - m(\pi)$.
It is parameterized using the Monte Carlo simulation.
The combinatorial background is described by an exponential function.
The $\jp K$, $\jp \pi$, and $\jp K^*$ contributions 
depend on the kaon momentum. The Monte Carlo simulation
shows that this dependence can be modeled by the same polynomial
function with different scaling factors for
$\jp K$, $\jp \pi$, and $\jp K^*$ signals.
The coefficients of the polynomial are determined
from the fit.
The $B \to \jpsi K$ signal contains $40,222 \pm 242$(stat.) events,
while the $B \to \jpsi \pi$ signal contains $1,578 \pm 119$(stat.) events.

To measure the charge asymmetry $A$ 
between the $\jp K^-(\pi^-)$ 
and $\jp K^+(\pi^+)$ final states, 
both physics and detector effects contributing to the
possible imbalance of events with positive and negative kaons must 
be taken into account. One physics source of asymmetry is direct CP violation
in the $B^+ \to \jp K^+(\pi^+)$ decay. 
In addition, forward-backward charge asymmetry
of events produced in the proton-antiproton collisions can also be present.
Detector effects can give rise to an artificial asymmetry if, for example, 
the reconstruction efficiencies of positive and negative particles 
are different. However, a positive particle produces the same track 
as a negative particle in the detector
with reversed magnet polarity. Therefore, essentially all detector effects 
can be canceled by regularly reversing the magnet polarity.

Following the method applied in \rf{bruce,asl},
the event sample of \fig{mjpk-fit} is divided into eight subsamples 
corresponding to all possible combinations of the solenoid polarity 
$\beta = \pm 1$, 
the sign of the  pseudorapidity of the $\jp K$ system $\gamma = \pm 1$, 
and the sign of the kaon candidate charge $q = \pm 1$.
In each subsample, the number $n^{\beta \gamma}_q$ of the events 
in the contributing channels, $\jp K$, $\jp \pi$ and $\jp K^*$, 
is obtained from the unbinned likelihood fit to the
mass distribution $m(\jp K)$ using the same likelihood function as 
for the whole sample.
All parameters of the fits apart from the fractions of the $\jp K$ signal,
the $\jp \pi$ signal, and the $\jp K^*$ signal, 
are fixed to the values determined from the fit to the whole sample.

The number of events in the $\jp K$ and $\jp \pi$ channels 
for each $\beta \gamma q$ subsample are
used to disentangle the physics asymmetries and the detector
effects. The $n^{\beta \gamma}_q$ can be expressed through the physics 
and the detector asymmetries as follows \cite{bruce}:
\begin{eqnarray}
n_{q}^{\beta \gamma} & = & \frac{1}{4} N \epsilon^{\beta}(1+qA)
(1+q \gamma A_{fb})
(1+\gamma A_{det}) \nonumber \\
                     & \times & (1+q \beta \gamma A_{q \beta \gamma}) 
(1+q \beta A_{q \beta})(1+\beta \gamma A_{\beta\gamma}).
\label{detpar}
\end{eqnarray}
Here
$N$                   is the total number of signal events;
$\epsilon^{\beta}$    is the fraction of integrated luminosity with solenoid polarity $\beta$ 
                      ($\epsilon^+ + \epsilon^- = 1$);
$A$                   is the charge asymmetry to be measured;
$A_{fb}$              accounts for possible forward-backward asymmetric $B$ meson production;
$A_{det}$             is the detector asymmetry for kaons emitted in the forward and backward direction;
$A_{q \beta \gamma}$  accounts for the change in acceptance of kaons of different sign
                      bent by the solenoid in different directions;
$A_{q \beta}$         is the detector asymmetry, which accounts for the change in
                      the kaon reconstruction efficiency when the solenoid polarity is reversed;
$A_{\beta \gamma}$    accounts for any detector-related forward-backward asymmetries 
                      that remain after the solenoid polarity flip.
We apply a $\chi^2$ fit of Eq.~\ref{detpar} to the number of events 
in all subsamples and extract all asymmetries and the total number of events 
in the $\jp K$ and $\jp \pi$ channels 
together with the fraction of events with 
positive solenoid polarity $\epsilon^+$, which is constrained to be
the same for both channels.
Results are presented in \tab{asym}. 
The charge asymmetry between $B^- \to \jp K^-$ and $B^+ \to \jp K^+$ 
is measured to be $A(\jp K)=-0.0070 \pm 0.0060$, and
the charge asymmetry between $B^- \to \jp \pi^-$ and $B^+ \to \jp \pi^+$ 
is found to be $A(\jp \pi)=-0.09 \pm 0.08$.
The detector asymmetries are all consistent with zero, since the acceptance
of the charged particles of different sign inside 
the solenoid is the same. However, 
we measure these asymmetries directly and do not rely on 
assumptions. The forward-backward asymmetry is also consistent with zero,
as expected in the SM.

\begin{table}
\caption{Physics and 
detector asymmetries for $J/\psi K$ and $J/\psi \pi$ channels.
$\epsilon^+$ is constrained to be the same for
both channels.}
\begin{ruledtabular}
\begin{tabular}{cD{+}{\,\pm\,}{-1}D{+}{\,\pm\,}{-1}}

&\multicolumn{1}{c}{$\jp K$}  & \multicolumn{1}{c}{$\jp \pi$} \\
\hline
$N$ & 40,217 + 243 & 1,577 + 118 \\
$\epsilon^{+}$ & \multicolumn{2}{c}{$0.5060 \pm 0.0030$}  \\
\hline
$      A$           & -0.0070 + 0.0060  & -0.09 + 0.08 \\
$ A_{fb}$           &  0.0013 + 0.0060  &  0.04 + 0.09 \\
$A_{det}$           & -0.0033 + 0.0060  &  0.21 + 0.08 \\
$A_{q\beta\gamma}$  & -0.0050 + 0.0060  & -0.02 + 0.09 \\
$A_{q\beta}$        &  0.0001 + 0.0060  & -0.19 + 0.08 \\
$A_{\beta\gamma}$   & -0.0030 + 0.0060  &  0.05 + 0.08 \\
\end{tabular}
\end{ruledtabular}
\label{asym}
\end{table}





\begin{figure}
\centering
\includegraphics[width=0.45\textwidth]{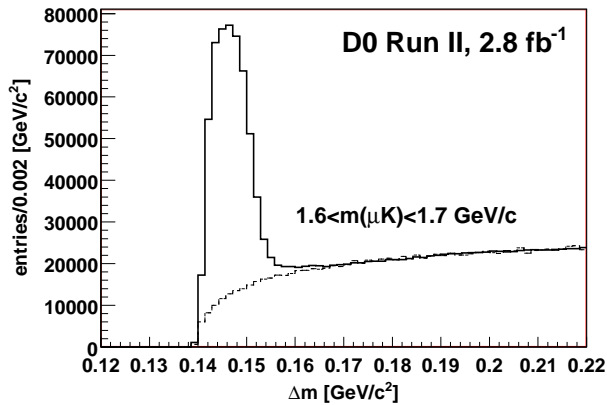}
\caption{The $\Delta m $ distribution (solid line) for
$1.6 < m(\mu K) < 1.7$ GeV/$c^2$.
The background distribution of events with wrong charge correlation 
of a muon and a kaon (dashed line) is also shown. The background normalization 
is required to give the same number of events with right and wrong charge
correlation for $0.19 < \Delta m < 0.22$ GeV/$c^2$.
}
\label{mmuk}
\end{figure}

In addition to the detector effects, the charge asymmetry 
$A(B \to \jp K)$ is affected by the difference in the
interaction cross-section of $K^+$ and $K^-$ with 
the detector material \cite{pdg}, which is
due to the fact that the reaction 
$K^- N \to Y \pi$ (where Y are hyperons $\Lambda$, $\Sigma$ etc.)
has no $K^+ N$ analog.
The difference in the interaction cross section results in a 
lower reconstruction
efficiency of $K^-$ and a visible kaon charge asymmetry $A_K$
between $K^-$ and $K^+$ candidates, which shifts the $A(\jp K)$ asymmetry.
The kaon asymmetry is measured directly in data
by comparing the exclusive decay
$c \to D^{*+} \to D^0 \pi^+$, $D^0 \to \mu^+ \nu_{\mu} K^-$
and its charge conjugate. 
It is expected from theory that there is no
CP violation in the semileptonic 
$D^0$ decays \cite{petrov}.
The possible CP-violating effects in $B \to D^{*\pm}X$
decays are estimated to give a negligible contribution.
Therefore, the observed asymmetry is only due to 
kaon reconstruction.
The decay of $D^*$ produces a clear peak in the mass difference,
$\Delta m = m(\mu K \pi) - m(\mu K)$. Its width depends on the mass $m(\mu K)$.
An example of the $\Delta m$ distribution for  $1.6 < m(\mu K) < 1.7$ GeV/$c^2$
is shown in Fig.~\ref{mmuk}. The combinatorial 
background under the peak is determined using events 
where all three particles (muon, kaon, and pion) have the same charge,
and its normalization is obtained using events with large values 
of $\Delta m$ outside the $D^*$ peak. The number of $D^* \to D^0 \pi$
decays is determined by subtracting the normalized number of background
events from the number of signal events in the mass band corresponding 
to the $D^*$ peak. The width of this band is varied depending on the mass
of the $\mu K$ system to ensure maximal signal significance.



\begin{figure}
\begin{center}
\includegraphics[width=0.45\textwidth]{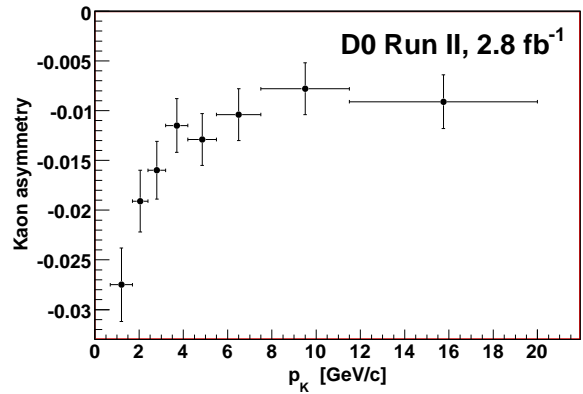}
\caption{Dependence of the kaon asymmetry on the kaon momentum
$p_K$ in eight bins 
of approximately equal statistics. Errors are statistical.}
\label{kasym-in-dst}
\end{center}
\end{figure}

The detector charge asymmetries
are disentangled from the kaon asymmetry using the same
detector model of Eq.~\ref{detpar}. To account for the momentum dependence
of the kaon cross-section \cite{pdg}, the kaon asymmetry is measured
in different bins of kaon momentum $p_K$, as shown in \fig{kasym-in-dst}.
The obtained asymmetry is convoluted with the kaon momentum distribution
in the $B \to \jp K$ decay and the resulting kaon asymmetry in the 
$B \to \jp K$ decay is found to be $A_K=-0.0145 \pm 0.0010$.
Taking into account this value, we obtain
$\adcpv=A(\jp K)-A_K=+0.0075 \pm 0.0061$(stat.)

The systematic uncertainty of  $\acp(B^+ \to \jp K^+)$ 
is estimated as follows.
The systematic uncertainty from the unbinned fit 
of the invariant mass distribution of the $\jp K$ system
is estimated by varying the parameters fixed during 
the unbinned fitting in the  $\beta \gamma q$
subsamples by $\pm 1 \sigma$, and is found to be 0.0002.
The systematic uncertainty from the choice of the fitting range is 
found to be 0.0004.
The shape of the  $\jp K^*$ contribution 
to the likelihood function is parameterized using the Monte Carlo simulation, 
and therefore produces an uncertainty in the number of signal events. 
We repeat the fit with different models of $\jp K^*$ contribution, including
a model without any such contribution. The maximal deviation 
in the resulting asymmetry is found to be 0.0025, which is taken as the 
systematic uncertainty from this source.
To measure the kaon asymmetry in the detector,
we subtract the combinatorial
background under the $D^*$ peak (see \fig{mmuk}, dashed line).
To estimate the uncertainty from the background definition,
we required the muon and the pion to have different charges and repeated
the measurement of the kaon asymmetry.
The resulting deviation in $\acp(B^+ \to \jp K^+)$ is found to be 0.0008.
Also, the sample used to measure the kaon asymmetry contains a contribution of 
$D^0$ semileptonic decays without a charged kaon in the final state. 
They are taken into account assuming the same selection efficiency as the
dominant $D^0 \to \mu \nu K$ decay. To find the impact of this assumption
on the final result, we repeated the measurement
of the kaon asymmetry assuming a zero reconstruction efficiency for additional
$D^0$ decay modes. The resulting deviation in $\acp(B^+ \to \jp K^+)$ 
is 0.0005. Combining all contributions in quadrature, 
we estimate the total systematic uncertainty on 
$\acp(B^+ \to \jp K^+)$ to be 0.0027,
which is dominated by the uncertainty from the
mass distribution model.


The systematic uncertainty of  $\acp(B^+ \to \jp \pi^+)$ is estimated 
similarly to that of $\acp(B^+ \to \jp K^+)$.
The only sizable contributions are:
0.01 from the variation of the fitting range,
and 0.02 from the mass model.
The total systematic uncertainty is 0.03.



In conclusion, the direct CP violating asymmetry in the $B^+ \to \jp K^+$ decay
is measured to be $\adcpv = +0.0075 \pm 0.0061$(stat.)$\pm0.0027$(syst.),
which is consistent with the world average, 
$A_{CP}(B^+ \to \jp K^+)=+0.015 \pm 0.017$ \cite{pdg},
but has  a factor of two improvement in precision, thus providing
the most stringent bounds for new models predicting
large values of $\adcpv$.
The direct CP violating asymmetry in  the $B^+ \to \jp \pi^+$ decay
is measured to be
$A_{CP}(B^+ \to \jp \pi^+) = -0.09 \pm 0.08$(stat.)$\pm 0.03$(syst.).
Our result agrees with the
previous measurements of this asymmetry \cite{pdg} and has a competitive
precision.

%
We thank the staffs at Fermilab and collaborating institutions, 
and acknowledge support from the 
DOE and NSF (USA);
CEA and CNRS/IN2P3 (France);
FASI, Rosatom and RFBR (Russia);
CNPq, FAPERJ, FAPESP and FUNDUNESP (Brazil);
DAE and DST (India);
Colciencias (Colombia);
CONACyT (Mexico);
KRF and KOSEF (Korea);
CONICET and UBACyT (Argentina);
FOM (The Netherlands);
STFC (United Kingdom);
MSMT and GACR (Czech Republic);
CRC Program, CFI, NSERC and WestGrid Project (Canada);
BMBF and DFG (Germany);
SFI (Ireland);
The Swedish Research Council (Sweden);
CAS and CNSF (China);
and the
Alexander von Humboldt Foundation.
%

\end{document}